\documentclass[pra,twocolumn,showpacs,superscriptaddress,floatfix]{revtex4}  
\usepackage{graphicx}
\usepackage{amssymb,amsmath,latexsym,amsthm}
\usepackage{bm}
\usepackage[dvips]{epsfig}
\usepackage{verbatim}
\usepackage{txfonts}

\renewcommand{\vec}[1]{\bm{\mathrm{#1}}}
\newcommand{\vhat}[1]{\hat{\bm{\mathrm{#1}}}}
\newcommand{\ocite}[1]{Ref.~[\onlinecite{#1}]}

\begin{document}

\title{Perpendicular magnetic anisotropy of two-dimensional Rashba ferromagnets}
\author{Kyoung-Whan Kim}%
\affiliation{Center for Nanoscale Science and Technology, National Institute
of Standards and Technology, Gaithersburg, Maryland 20899, USA}%
\affiliation{Maryland NanoCenter, University of Maryland, College Park,
Maryland 20742, USA}%
\affiliation{Basic Science Research Institute, Pohang University of
Science and Technology, Pohang 37673, Korea}%
\affiliation{PCTP and Department of Physics, Pohang University of Science and
Technology, Pohang 37673, Korea}%
\author{Kyung-Jin Lee}%
\affiliation{Department of Materials Science and Engineering, Korea
University, Seoul 02841, Korea}%
\affiliation{KU-KIST Graduate School of Converging Science and Technology,
Korea University, Seoul 02841, Korea}%
%
\author{Hyun-Woo Lee}
\email{hwl@postech.ac.kr}%
\affiliation{PCTP and Department of Physics, Pohang University of Science and
Technology, Pohang 37673, Korea}%
\author{M. D. Stiles}%
\email{mark.stiles@nist.gov}%
\affiliation{Center for Nanoscale Science and Technology, National Institute
of Standards and Technology, Gaithersburg, Maryland 20899, USA}%
\date{\today}
%

\begin{abstract}
We compute the magnetocrystalline anisotropy energy within two-dimensional
Rashba models. For a ferromagnetic free-electron Rashba model, the magnetic anisotropy is exactly zero regardless of the strength of the Rashba coupling, unless only the lowest band is occupied. For this latter case, the model predicts in-plane anisotropy. For a more realistic Rashba model with finite band width, the magnetic anisotropy evolves from in-plane to perpendicular and back to in-plane as bands are progressively filled. This evolution agrees with first-principles calculations on the interfacial anisotropy, suggesting that the Rashba model captures energetics leading to anisotropy originating from the interface provided that the model takes account of the finite Brillouin zone. The results show that the electron density modulation by doping or an external voltage is more important for voltage-controlled magnetic anisotropy than the modulation of the Rashba parameter.
\end{abstract}

\pacs{}

\maketitle

\section{Introduction \label{Sec:Intro}}

Recent developments in the design of spintronic devices favor perpendicular
magnetization, increasing the interest in materials with perpendicular
magnetic anisotropy~\cite{Carcia85APL,Draaisma87JMMM,Monso02APL,Ikeda10NM}.
One advantage is that devices with the same thermal stability can be switched more easily if the magnetization is perpendicular than if it is in plane~\cite{Ikeda10NM,Jung08APL,Nakayama08JAP,Sbiaa11JAP,Heinonen10JAP,Worledge11APL}.
Since magnetostatic interactions favor in-plane magnetization for a thin film geometry, perpendicular magnetic anisotropy requires materials and interfaces that have strong magnetocrystalline anisotropy. Numerous computational studies~\cite{Daalderop92PRL,Daalderop94PRB,Sakuma94JPSJ,Nakamura10PRB,Yang11PRB,Odkhuu13PRB,Khoo13PRB,Hallal13PRB}
show the importance of interfaces on magnetocrystalline anisotropy. The theory developed by Bruno~\cite{Bruno89APA,Bruno89PRB}, which provides an insightful explanation of the surface magnetocrystalline anisotropy and its correlation with orbital moment~\cite{Oppeneer98JMMM}, has been confirmed by experiments~\cite{Weller95PRL,Shaw13PRB}. The cases for which the Bruno's theory does not apply~\cite{Andersson07PRL} require a case by case study through first-principles calculations, making it difficult to get much insight.

Some insight into perpendicular magnetic anisotropy can be gained by studying it within a simple model. One such model is the two-dimensional Rashba model~\cite{Bychkov84JETPL}.  A two-dimensional Rashba model includes only minimal terms imposed by symmetry breaking. As extensive theoretical studies have shown, a two-dimensional Rashba model can capture most of the qualitative physics of spin-orbit coupling with broken inversion symmetry, such as the intrinsic spin Hall
effect~\cite{Sinova04PRL,Inoue04PRB}, the intrinsic anomalous Hall
effect~\cite{Inoue06PRL}, the fieldlike spin-orbit
torque~\cite{Manchon08PRB,Matos-A09PRB}, the dampinglike spin-orbit
torque~\cite{Wang12PRL,Kim12PRB,Pesin12PRB,Kurebayashi14NN}, the
Dzyaloshinskii-Moriya interaction~\cite{Dzyaloshinskii57SPJ,Moriya60PR,Fert80PRL,Kim13PRL}, chiral spin motive forces~\cite{Kim12PRL,Tatara13PRB}, and corrections to the
magnetic damping~\cite{Kim12PRL}, each of which has received attention because of its relevance for efficient device applications. Despite the extensive studies, exploring magnetocrystalline anisotropy within the simple model is still limited. Magnetocrystalline anisotropy derived from a two-dimensional Rashba model may clarify the correlations between it and various physical quantities listed above.

There are recent theoretical and experimental studies on the possible correlation between the magnetic anisotropy and the Rashba spin-orbit coupling strength. The theories~\cite{Xu12JAP,Barnes14SR} report a simple proportionality relation between perpendicular magnetic anisotropy and square of the Rashba spin-orbit coupling strength and argued its connection to the voltage-controlled magnetic anisotropy~\cite{Nakamura10PRB,Weisheit07Science,Duan08PRL,Maruyama09NN,Wang12NM,Shiota13APL}. However, these experiments require further interpretation. Nistor \emph{et al.}~\cite{Nistor10IEEE} report the positive correlation between the Rashba spin-orbit coupling strength and the perpendicular magnetic anisotropy while Kim \emph{et al.}~\cite{Kim16APL} report an enhanced perpendicular magnetic anisotropy accompanied by a reduced Dzyaloshinskii-Moriya interaction in case of Ir/Co. Considering that the Dzyaloshinskii-Moriya interaction and the Rashba spin-orbit coupling are correlated according to \ocite{Kim13PRL}, the perpendicular magnetic anisotropy and the Rashba spin-orbit coupling vary opposite ways in the latter experiment. These inconsistent observations imply that the correlation is, even if it exists, not a simple proportionality. In such conceptually confusing situations, simple models, like that in this work, may provide insight into such complicated behavior.

In this paper, we compute the magnetocrystalline anisotropy within a
two-dimensional Rashba model in order to explore the correlation between the magnetocryatalline anisotropy and the Rashba spin-orbit coupling. We start from Rashba models added to different kinetic dispersions (Sec.~\ref{Sec:Model}) and demonstrate the following core results. First, a two-dimensional ferromagnetic Rashba model with a free electron dispersion results in \emph{exactly} zero anisotropy  once the Fermi level is above a certain threshold value (Sec.~\ref{Sec:Result-A}). This behavior suggests that the simple model is not suitable for studying the magnetic anisotropic energy in that regime. Second, simple modifications of the model do give a finite magnetocrystalline anisotropy proportional to the square of the Rashba parameter (Sec.~\ref{Sec:Result-B}). We illustrate with tight-binding Hamiltonians that a Rashba system acquires perpendicular magnetic anisotropy for wide parameter ranges once the Brillouin zone and energy band width being finite in size is taken into account in the model. This demonstrates that the absence of magnetic anisotropy is a peculiar feature of the former free-electron Rashba model and we discuss the similarity of this behavior to the intrinsic spin Hall conductivity~\cite{Inoue04PRB}. Third, we show that the magnetocrystalline anisotropy of the modified Rashba models strongly depends on the band filling (Sec.~\ref{Sec:Result-B}). The system has in-plane magnetic anisotropy for low band filling. As the electronic states are occupied, the anisotropy evolves from in-plane to perpendicular and back to in-plane for high electron density. This suggests that it may be possible to see such behavior in systems in which the interfacial charge density can be modified, for example through a gate voltage. This also provides a way to reconcile mutually contradictory experimental results~\cite{Nistor10IEEE,Kim16APL} since different band filling can result in opposite behaviors of the magnetocrystalline anisotropy. We make further remarks in Sec.~\ref{Sec:Result-C} and summarize the paper in Sec.~\ref{Sec:Conclusion}. We present the analytic details in Appendix.

\section{Model and formalism\label{Sec:Model}}

We first present the model and formalism for a quadratic dispersion and then generalize the model to a tight-binding dispersion. In this paper, we call a Rashba model with a quadratic dispersion a ``free-electron Rashba model" and call a Rashba model with a tight-binding dispersion a ``tight-binding Rashba model". All the models include ferromagnetism in the same manner.

\begin{figure}
\includegraphics[width=6.4cm]{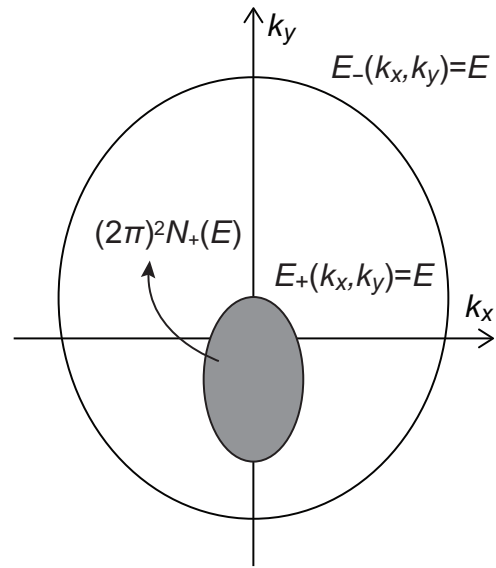}
\caption{Geometrical meaning of $N_+(E)$, the number of minority electrons per unit area that satisfies $E_+(k_x,k_y)\le E$. $N_+(E)$
is given by the area enclosed by the constant energy contour of $E_+(k_x,k_y)=E$.
$N_-(E)$, the number of majority electrons per unit area that satisfies $E_-(k_x,k_y)\le E$, has the similar meaning (not shown in the figure).}
\label{Fig:electron_density}
\end{figure}

A ferromagnetic free-electron Rashba model is described by the following Hamiltonian.
\begin{equation}
H=\frac{\vec{p}^2}{2m_e}+J\vec{\sigma}\cdot\vec{m}+\frac{\alpha_R}{\hbar}(\vec{\sigma}\times\vec{p})\cdot\vhat{z},\label{Eq:Rashba_quadratic}
\end{equation}
where $\vec{p}$ is the momentum operator of itinerant electrons, $m_e$ is the effective electron mass, $J>0$ is the exchange energy between conduction
electrons and the magnetization, $\vec{\sigma}$ is the vector of the Pauli spin matrices,
$\alpha_R$ is the Rashba parameter, $\vhat{z}$ is the interface normal
direction perpendicular to the two-dimensional space, and $\vec{m}$ is a unit vector along the direction of magnetization. The terms in Eq.~(\ref{Eq:Rashba_quadratic}) reflect the quadratic kinetic energy, the exchange interaction, and the Rashba spin-orbit coupling, respectively. The second and third term originate respectively from the time-reversal symmetry breaking (magnetism) and the space-inversion symmetry breaking (interface). Thus, the Rashba model is a minimal model taking account of the symmetry breaking features of the system. There are various types of Rashba models depending on the momentum dependence of spin-orbit coupling Hamiltonian~\cite{Gerchikov92SPS}. We confine the scope of the paper to the linear Rashba model that is linear in $\vec{p}$ [the last term in Eq.~(\ref{Eq:Rashba_quadratic})] and is the most widely used form. We emphasize that the Rashba model is mainly useful for its pedagogical value rather than its ability to make quantitative predictions for real materials~\cite{Krasovskii14PRB,Grytsyuk16PRB}. In Ref.~\cite{Grytsyuk16PRB}, the authors find that while it is possible to extract an effective Rashba parameter for realistic interfaces, it was not possible to connect this parameter to the calculated magnetocrystalline anisotropy. Still, even though the simple Rashba model may have only limited direct connection to the electronic structure of most interfaces of interest, it does provide a qualitative understanding of their physical properties.

Diagonalization of Eq.~(\ref{Eq:Rashba_quadratic}) gives the single particle
energy spectrum of the free-electron Rashba model. For a homogeneous magnetic texture, $H$ commutes with $\vec{p}$, thus $\vec{k}=\vec{p}/\hbar$ is a good quantum number. In terms of $\vec{k}$, diagonalization of the $2\times2$ Hamiltonian gives the energy eigenvalues $E_\pm(k_x,k_y)$ of $H$ for spin majority and minority bands, where $+$ and $-$ refer to minority and majority bands respectively.
\begin{equation}
E_\pm(k_x,k_y)=\frac{\hbar^2k^2}{2m_e}\pm\sqrt{J^2+2J\alpha_R(k_ym_x-k_xm_y)+\alpha_R^2k^2},\label{Eq:quadratic E_pm}
\end{equation}
where $k=|\vec{k}|$. Since the system has rotational symmetry around
$\vhat{z}$ axis~\cite{Garello13NN}, we assume $m_y=0$ from now on.

The total electron energy is given by summing up single particle energies at all electronic states below the Fermi level. To do this, we define $N_\pm(E)$, the number of minority/majority electrons per unit area that satisfies $E_\pm(k_x,k_y)\le E$. The geometrical meaning of $N_\pm(E)$ is the area enclosed by the constant energy contour $E_\pm(k_x,k_y)=E$ (Fig.~\ref{Fig:electron_density}). With this definition, the density of states for each band is given by $dN_\pm/dE$. Therefore, the expression of the total energy per unit area is given by
\begin{equation}
E_{\rm tot}(E_F)=\int_{E_{\rm min}^-}^{E_F}E\frac{dN_-}{dE}dE+\eta\int_{E_{\rm min}^+}^{E_F}E\frac{dN_+}{dE}dE,\label{Eq:total energy}
\end{equation}
where $E_{\rm min}^\pm$ is the band bottom energy of each band, below which
$N_\pm(E)=0$. $\eta=0$ if $E_F<E_{\rm min}^+$ so that there is no occupied
minority state, and $\eta=1$ otherwise. Such a factor is absent for the first term because we only consider the Fermi level $E_F$ above $E_{\rm min}^-$. Otherwise, the magnetocrystalline anisotropy is trivially zero since there is no occupied state. The total energy density depends on the direction of magnetization in general. We then compute the magnetocrystalline anisotropy by the difference of the total energy density for perpendicular and in-plane
magnetizations; $\Delta E=E_{\rm tot}|_{\vec{m}=\vhat{x}}-E_{\rm
tot}|_{\vec{m}=\vhat{z}}$.

To compute $\Delta E$ from Eq.~(\ref{Eq:total energy}), the Fermi levels for $\vec{m}=\vhat{x}$ and $\vec{m}=\vhat{z}$ need to be specified. Since the energy dispersion [Eq.~(\ref{Eq:quadratic E_pm})] is in general dependent on $\vec{m}$, the Fermi level also changes as a function of $\vec{m}$, because the total electron density does not change for an isolated magnetic system. Thus, we fix the total electron density as a constraint. To fix the total electron density as a constraint, we define the total electron density below energy $E$.
\begin{equation}
N_e(E)=N_-(E)+\eta N_+(E).\label{Eq:total number density}
\end{equation}
The domain of $E$ is $E\ge E_{\rm min}^-$ so that $N_e(E)\ge0$. Since $N_e(E)$ is a strictly increasing function of $E$ in the domain, it has an inverse function in $N_e\ge 0$. We denote the inverse function by $\varepsilon_F(N_e)$. $\varepsilon_F$ has the same
physical meaning as the Fermi level $E_F$ for a given electron density $N_e$. However, we use the different symbols to emphasize that $\varepsilon_F$ is given by a \emph{function} of the electron density while $E_F$ is just a
\emph{given constant}. With this definitions, the magnetocrystalline
anisotropy is given by
\begin{equation}
\Delta E(N_e)=E_{\rm
tot}\left(\varepsilon_F(N_e)\right)|_{\vec{m}=\vhat{x}}-E_{\rm tot}\left(\varepsilon_F(N_e)\right)|_{\vec{m}=\vhat{z}}.\label{Eq:fixed electron density}
\end{equation}
This is the central equation of the formalism to compute the magnetocrystalline anisotropy.

We now compute the magnetic anisotropy for a tight-binding Rashba model. To construct a tight-binding Hamiltonian, we discretize
Eq.~(\ref{Eq:Rashba_quadratic})~\cite{Mireles01PRB,Pareek02PRB}. In the main text, we use a tight-binding Hamiltonian for a two-dimensional square lattice as an example. The construction and the results of a tight-binding Hamiltonian for a two-dimensional hexagonal lattice (equivalently a triangular lattice) are presented in Appendix~\ref{Sec(A):hexagonal}. For simplicity, we use a two-band tight-binding Hamiltonian including spin degrees of freedom only, but ignoring all orbital degrees of freedom.
\begin{subequations}
The tight-binding Hamiltonian we construct here is given by
\begin{equation}
H=H_K+H_J+H_R,
\end{equation}
where $H_K$, $H_J$, and $H_R$ are the discretized versions of the kinetic energy, the exchange energy, and the Rashba Hamiltonian, respectively. $H_K$ is constructed by the hopping terms to the nearest neightbor sites.
\begin{equation}
H_K=-\frac{\hbar^2}{2m_ea^2}\sum_{pq\sigma}(C_{p+1,q,\sigma}^\dagger C_{p,q,\sigma}+C_{p,q+1,\sigma}^\dagger C_{p,q,\sigma})+\mathrm{h.c.},
\end{equation}
where $a$ is the lattice constant, $p$ and $q$ are the site indicies, and
$C_{p,q,\sigma}$ is the electron annihilation operator at site
$(x,y)=(pa,qa)$ with spin $\sigma$. h.c. refers to hermitian conjugate of all the terms in front of it. Each term in the summand corresponds to hopping to $x$ and $y$ directions respectively. The hopping parameter $-(\hbar^2/2m_ea^2)$ is determined by matching the energy dispersion with the free electron dispersion $\hbar^2k^2/2m_e$ in the continuum limit $a\to0$. $H_J$ is constructed by on-site energy that mixes the spin degree of freedom.
\begin{equation}
H_J=J\sum_{pq\sigma\sigma'}\left[C_{p,q,\sigma}^\dagger(\vec{\sigma})_{\sigma,\sigma'}C_{p,q,\sigma'}\right]\cdot\vec{m},
\end{equation}
where $(\vec{\sigma})_{\sigma,\sigma'}$ is the matrix element of the Pauli
matrices. $H_R$ is constructed as following. We impose a hopping term from a
site to a neighboring site, along a direction $\vhat{u}$. Since $\vhat{u}$ corresponds to the electron momentum direction, the term
acquires a spin Pauli matrix $(\vec{\sigma}\times\vhat{u})\cdot\vhat{z}$.
Then, a hopping term along the $y$ direction acquiring $\sigma_x$ is given by $itC_{p,q+1,\sigma'}^\dagger(\sigma_x)_{\sigma,\sigma'}C_{p,q,\sigma}$, where $t$ is a real hopping parameter. After considering all the neighboring hopping terms satisfying the hermiticity condition, we determine the hopping
parameter by taking continuum limit up to $\mathcal{O}(a^2)$ and matching the energy dispersion with Eq.~(\ref{Eq:quadratic E_pm}). In this way, we end up
with
\begin{align}
H_R&=i\frac{\alpha_R}{2a}\sum_{pq\sigma\sigma'}\left[C_{p,q+1,\sigma}^\dagger(\sigma_x)_{\sigma,\sigma'}C_{p,q,\sigma}\right.\nonumber\\
&\phantom{=i\frac{\alpha_R}{2a}\sum_{pq\sigma\sigma'}[~}\left.-C_{p+1,q,\sigma}^\dagger(\sigma_y)_{\sigma,\sigma'}C_{p,q,\sigma}\right]+\mathrm{h.c.}.
\end{align}
\label{Eq:TB Hamiltonian}
\end{subequations}
For more details of determining the hopping parameters, see the example in Appendix~\ref{Sec(A):hexagonal} for a two-dimensional hexagonal lattice.

\begin{figure}
\includegraphics[width=7.6cm]{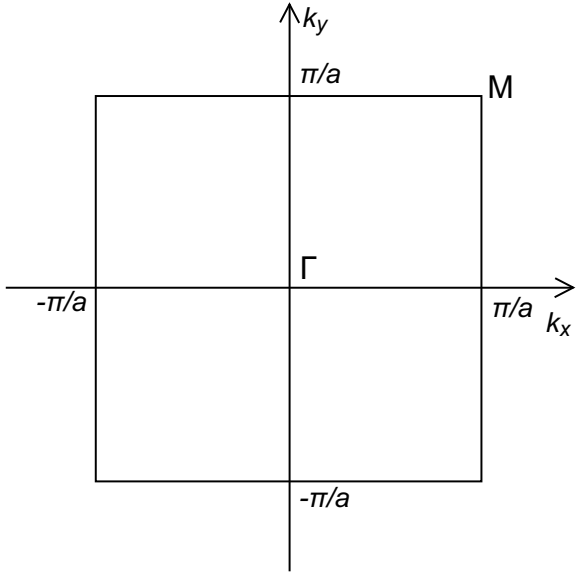}
\caption{Brillouin zone of Eq.~(\ref{Eq:TB Hamiltonian}).
We denote the $\rm \Gamma$ and $\rm M$ points for later purpose.
}
\label{Fig:BZ}
\end{figure}

Now we use the same formalism [Eq.~(\ref{Eq:fixed electron density})]. We use the discrete translational symmetry of the lattice to use the Bloch theorem and compute the energy dispersion relation as a function of the crystal momentum. One difference is that the Brillouin zone and the band width for a tight-binding Hamiltonian are finite (Fig.~\ref{Fig:BZ}), while these are infinite for the free electron model Eq.~(\ref{Eq:Rashba_quadratic}). Therefore, the domain of the integration in Eq.~(\ref{Eq:total energy}) is not only limited by the Fermi contour, but also limited by the Brillouin zone boundary. We show in Sec.~\ref{Sec:Result-B} that the finite band width is a
crucial feature for emergence of perpendicular magnetic anisotropy for wide ranges of parameters.

\section{Magnetocrystalline anisotropy\label{Sec:Result}}

\subsection{Free-electron Rashba model\label{Sec:Result-A}}

Although the free electron model we present above [Eq.~(\ref{Eq:Rashba_quadratic})] has a simple form, it still requires
complicated mathematics to assess the magnetocrystalline anisotropy predicted by the model since a constant energy contour given by
Eq.~(\ref{Eq:quadratic E_pm}) is a quartic curve. In this
section, we first discuss results of a perturbative analysis, which assumes $\alpha_R$ to be small and keeps terms only up to $\mathcal{O}(\alpha_R^2)$. In this regime, a constant energy contour is a quadratic curve which allows the magnetocrystalline anisotropy to be calculated analytically. The analytic results shall give useful insight into the model. We then go beyond the perturbative regime and discuss exact results in the nonperturbative regime with arbitrary $\alpha_R$. In particular, we check if the conclusions from the perturbative analysis remain valid in the nonperturbative regime.

\subsubsection{Perturbation theory: Insights into the model}

As we show in Appendix.~\ref{Sec:Detail-A}, expanding Eq.~(\ref{Eq:quadratic E_pm}) up to $\mathcal{O}(\alpha_R^2)$, we obtain a quadratic equation with respect to
$(k_x,k_y)$. The contour $E_\pm(k_x,k_y)=E$ forms an ellipse, by which the area
enclosed is exactly computable. Since $N_\pm(E)$ is exactly given in a simple
way, calculating Eq.~(\ref{Eq:total energy}) is straightforward. In this
perturbative regime, the relation between electron number density and the
Fermi level Eq.~(\ref{Eq:total number density}) is linearly given so
inverting $N_e(E)$ is also straightforward. Then, the magnetic anisotropy $\Delta E(N_e)$ [Eq.~(\ref{Eq:fixed electron density})] is evaluated after simple algebra.

There are two different regimes; $E_F<E_{\rm min}^+$ and $E_F\ge E_{\rm
min}^+$. For the first case, there are no minority electrons. For this case, $\eta=0$ in Eq.~(\ref{Eq:total energy}). In the second case, the minority band is also occupied. For this case, $\eta=1$ in Eq.~(\ref{Eq:total energy}). We examine the cases one by one.

When only majority band is occupied ($\eta=0$), the magnetocrystalline anisotropy [Eq.~(\ref{Eq:fixed electron density})] is
\begin{equation}
\Delta E(N_e)=-\frac{N_em_e\alpha_R^2}{2\hbar^2}\left(1-\frac{N_e}{N_-(E_{\rm min}^+)}\right)~~\mathrm{(majority~only)}.\label{Eq:MCA majority only}
\end{equation}
Here $N_-(E_{\rm min}^+)$ is the electron density when the Fermi level
touches the bottom of the minority band. The result shows that the magnetocrystalline anisotropy is at least quadratic in $\alpha_R$. Below we show this is a result of symmetry that the magnetocrystalline anisotropy should be an even function of $\alpha_R$. Equation~(\ref{Eq:MCA majority
only}) is valid only when there is no minority electrons $0<N_e<N_-(E_{\rm
min}^+)$. We show in Appendix~\ref{Sec:Detail-A} that $N_-(E_{\rm
min}^+)=Jm_e/\pi\hbar^2+\mathcal{O}(\alpha_R^2)$, which is independent of
$\vec{m}$~\footnote{We can ignore the $\mathcal{O}(\alpha_R^2)$ term here
since Eq.~(\ref{Eq:MCA majority only}) is already proportional to $\alpha_R^2$.}. Since $N_e<N_-(E_{\rm min}^+)$, Eq.~(\ref{Eq:MCA majority only}) predicts the magnetocryatalline anisotropy to be negative. The sign corresponds to in-plane magnetic anisotropy, which is counter to the na\"{i}ve expectation that the Rashba spin-orbit coupling generates the perpendicular magnetic anisotropy. However, this observation does not contradict experimental observations showing perpendicular magnetic anisotropy since experimental results are usually obtained when both bands are occupied.

Next we examine the second regime where both bands are occupied ($\eta=1$). Strikingly,
the same formalism leads us
\begin{equation}
\Delta E(N_e)=0~~\mathrm{(both~bands~occupied)},\label{Eq:absence MCA}
\end{equation}
regardless of $N_e$. There is no magnetocrystalline anisotropy for this case. An intuitive way to understand this striking behavior is observing the absence of angular dependence of $N_e$ as a function of the Fermi level. In Appendix~\ref{Sec:Detail-A}, we show that, once both bands are occupied,
\begin{equation}
(2\pi)^2 N_e(E_F)=\frac{4\pi m_e(m_e\alpha_R^2+E_F\hbar^2)}{\hbar^4},\label{Eq:Ne vs EF}
\end{equation}
which has no $\vec{m}$ dependence. Therefore, when we increase the number of electrons slightly by $dN_e$, the contribution to the additional magnetocrystalline anisotropy is $E_FdN_e=[(\pi \hbar^2N_e/m_e)-(m_e\alpha_R^2/\hbar^2)]dN_e$. Since this is independent of the direction of magnetization, adding electrons does not change the magnetocrystalline anisotropy at all. By noting that Eq.~(\ref{Eq:MCA majority only}) vanishes $N_e=N_-(E_{\rm min}^+)$, we end up with Eq.~(\ref{Eq:absence MCA}).

There is a recent theory~\cite{Barnes14SR} which predicts perpendicular
magnetic anisotropy with the free-electron Rashba model. In that work, the magnetocrystalline anisotropy is expressed by a characteristic energy denoted by $T$. Here we show that $T$ takes a value within that model such that the anisotropy is strictly zero.

\begin{figure}
\includegraphics[width=8.4cm]{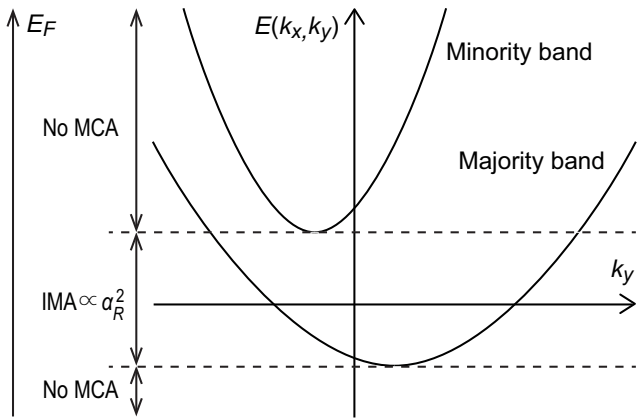}
\caption{Summary of the results of the magnetocrystalline anisotropy (MCA) from
the ferromagnetic free-electron Rashba model. The lower band and the upper band are respectively
majority and minority band. The horizonal and vertical displacements of each band
are respectively due to Rashba spin-orbit coupling and the exchange splitting.
The diagram shows behaviors of the magnetocrystalline anisotropy for each region.
The calculated magnetocrystalline anisotropy shows in-plane magnetic anisotropy (IMA)
when electrons in the ground state occupy only the majority band,
and the magnetic anisotropy energy is at least quadratic in the Rashba parameter $\alpha_R$.
On the other hand, the magnetocrystalline anisotropy vanishes once both bands
are partially occupied in the ground state.}
\label{Fig:summary_quadratic}
\end{figure}

To summarize this section, by using a perturbative approach, we make the following observations. First, the free-electron Rashba model model gives the magnetocrystalline anisotropy that is at least quadratic in $\alpha_R$.
Second, the model does not give perpendicular magnetic anisotropy. Third, the magnetocrystalline anisotropy vanishes unless only a single band is occupied. We summarize the result in Fig.~\ref{Fig:summary_quadratic}.

\subsubsection{Beyond perturbation: Extension of validity}

So far, we examined the properties of the free-electron Rashba model in the perturbative regime. The perturbative approach allows gaining insight into the model easily but it works only for small $\alpha_R$. In this section, we go beyond the perturbative regime to see if the conclusions we made in the previous section change when $\alpha_R$ is not small. We prove that the qualitative results from the perturbative analysis remain valid for large $\alpha_R$ as well.

\begin{figure}
\includegraphics[width=8.6cm]{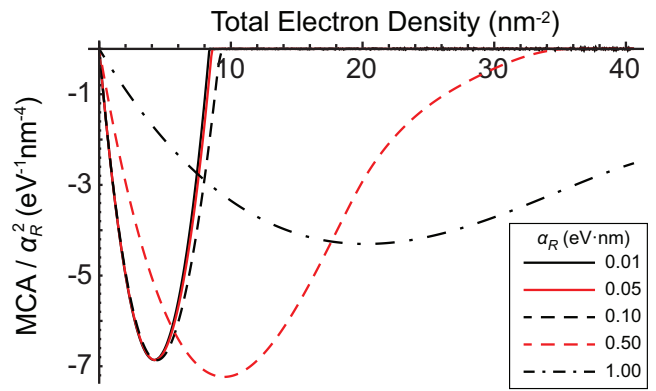}
\caption{(color online) Numerical computation of the magnetocryatalline anisotropy (MCA) divided by $\alpha_R^2$ within the free-electron Rashba model.
The results show in-plane magnetic anisotropy for wide range of the Rashba parameter and zero anistropy
after a certain threshold within the numerical error. Note that the magnetocrystalline anisotropy
is proportional to $\alpha_R^2$ for small Rashba parameters, confirming the result of our perturbation approach.
We use $J=1~\mathrm{eV}$ for the simulation.} \label{Fig:quadratic_numerical}
\end{figure}

First we prove that the magnetocrystalline anisotropy is at least quadratic in $\alpha_R$. For this, we consider the sign reversal of $\alpha_R$. This does not affect the energy eigenvalue spectrum of the Hamiltonian at all since the energy eigenvalue satisfies the property, $E(k_x,k_y;\alpha_R)=E(-k_x,-k_y;-\alpha_R)$ [see Eq.~(\ref{Eq:quadratic E_pm})]. Since the total energy density cannot change by a rotational transformation, it should be invariant under $\alpha_R\to-\alpha_R$. Therefore, the magnetocrystalline anisotropy may be expanded as a power series of $\alpha_R^2$ with the leading order term proportional to $\alpha_R^2$~\footnote{By the same transformation, we also end up with that the total energy density is expanded by $(1-m_z^2)$.}. When $\alpha_R$ becomes larger, higher order terms in $\alpha_R^2$ can contribute. In Fig.~\ref{Fig:quadratic_numerical}, we numerically compute the
magnetocrystalline anisotropy divided by $\alpha_R^2$. We see that the first
three curves almost overlap with each other. However, when $\alpha_R$ becomes larger so $\alpha_R k_F$ is comparable to $J$, the magnetocryatalline anisotropy divided by $\alpha_R^2$ varies as $\alpha_R$ changes, implying the breakdown of the perturbative result [Eq.~(\ref{Eq:MCA majority only})].

Although the perturbation theory breaks down quantitatively, qualitative
features remain the same for a wide range of $\alpha_R$. In particular, Fig.~\ref{Fig:quadratic_numerical} shows that the magnetocrystalline anisotropy predicted by the free-electron Rashba model is negative (in-plane magnetic anisotropy) for low electron density and vanishes (within the numerical error of our calculation) once the total electron density goes above threshold value. Perpendicular magnetic anisotropy is never generated.

It turns out that Eq.~(\ref{Eq:absence MCA}) can be rigourously proven for arbitrary $\alpha_R$. Due to its complexity, here we sketch the proof only briefly. The detailed proof is presented in Appendix~\ref{Sec:Detail-B}. The proof proceeds as follows. First, we consider a situation where both bands are occupied for both $\vec{m}=\vhat{x}$ and
$\vec{m}=\vhat{z}$, which occurs if and only if $E_F\ge
J$~\footnote{It is still possible for the minority band to be occupied for
$\vec{m}=\vhat{x}$ but not for $\vec{m}=\vhat{z}$. This case is not eligible for the proof.
The assumption requires that the minority band should be occupied regardless
of the direction of magnetization. For more information, see
Appendix~\ref{Sec:Detail-B}.}. We then use the Cauchy integral formalism for
complex contour integrals to show that Eq.~(\ref{Eq:Ne vs EF}) holds beyond the perturbative regime. As discussed in the previous section, Eq.~(\ref{Eq:Ne vs EF}) implies that the magnetocrystalline anisotropy is independent of the Fermi level when $E_F\ge J$. Next, we show that $\Delta E$ vanishes in the large $E_F$ limit. When combined together these features prove that $\Delta E$ should be exactly zero for $E_F\ge J$, which is nothing but Eq.~(\ref{Eq:absence MCA}).

Here we emphasize that although Eq.~(\ref{Eq:absence MCA}) holds for arbitrary $\alpha_R$, it is very unstable with respect to the model variation since it is crucially dependent on $N_e$ being independent of the magnetization $\vec{m}$ [Eq.~(\ref{Eq:Ne vs EF})], which holds only for the idealized free-electron Rashba model [Eq.~(\ref{Eq:Rashba_quadratic})]. Various types of modification of the Rashba model which make it more realistic can break this independence and result in the violation of Eq.~(\ref{Eq:absence MCA}). Possible deformations include the change of dispersion from strictly quadratic and truncation of the infinite band width to finite width. In the next section, we consider a tight-binding Rashba model, which is more realistic than the idealized free-electron Rashba model in the sense that the former has finite band width whereas the latter has infinite band width. This model shows that Eq.~(\ref{Eq:absence MCA}) is indeed violated and perpendicular magnetic anisotropy emerges. In passing, we note that not only the magnetocrystalline anisotropy but also other properties of the idealized free-electron Rashba model are peculiar. A well known example is the intrinsic spin Hall conductivity~\cite{Sinova04PRL,Inoue04PRB}. For the idealized free-electron Rashba model, it vanishes identically when both bands are partially filled but for slightly modified Rashba models~\cite{Murakami04PRB,Nomura05PRB}, it is finite.

\subsection{Tight-binding Rashba model\label{Sec:Result-B}}

\begin{figure}
\includegraphics[width=8.6cm]{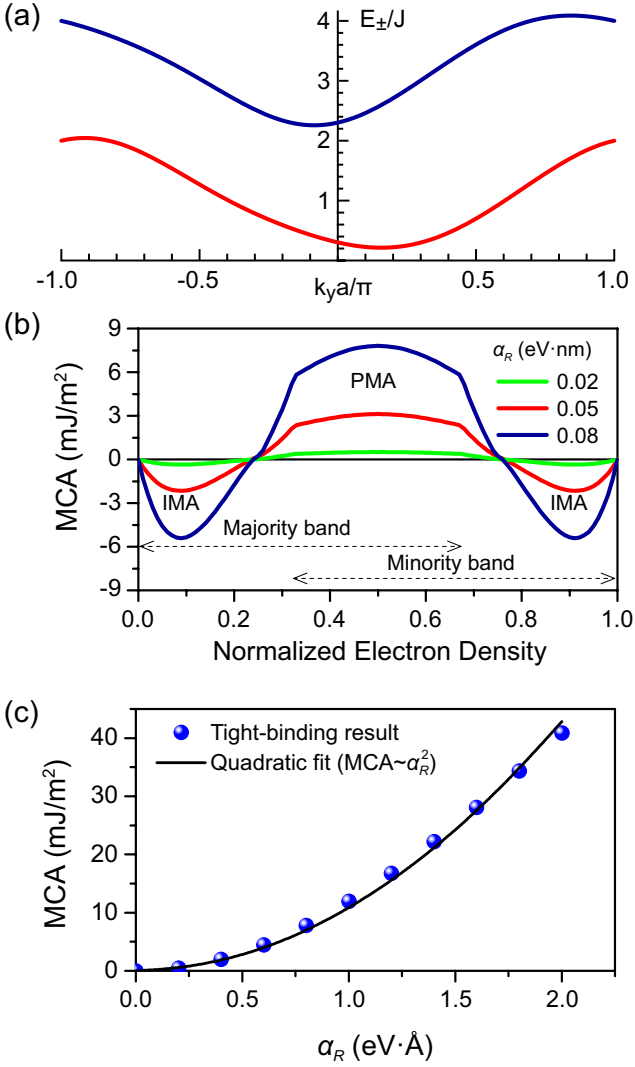}
\caption{(color online) (a) Energy dispersion of the tight-binding Rashba model [Eq.~(\ref{Eq:energy dispersion - tight binding})]. Energies as a function of $k_y$, for $k_x=0$, are given by the red and blue curves for majority and minority bands respectively. The asymmetry between $-k_y$ and $k_y$ originates from the Rashba interaction which is taken to be $\alpha_R=0.02~\mathrm{eV}\cdot\mathrm{nm}$ in this figure. Near the band minimum, the energy dispersion can be approximated by a quadratic dispersion [Eq.~(\ref{Eq:quadratic E_pm})]. However, near the Brillouin zone boundary, it differs significantly. The magnetization direction is taken to be $\vec{m}=(1/2,0,\sqrt{3}/2)$. Depending on the Rashba parameter and the direction of magnetization, the dispersion can be even more complicated particularly when the two bands approach each other. (b) The magnetocrystalline anisotropy (MCA)
as a function of the total electron density divided by the
electron density for completely filled bands $N_{\rm max}=2.2\times10^{19}\rm m^{-2}$. We use the Rashba parameters
$\alpha_R=0.02~\mathrm{eV}\cdot$nm,$~0.05~\mathrm{eV}\cdot$nm, and $0.08~\mathrm{eV}\cdot$nm. The results show in-plane magnetic
anisotropy (IMA) for very low and very
high electron occupation. For a wide range of the intermediate electron density, it shows
perpendicular magneitc anisotropy (PMA). (c) The peak values of the
magnetocrystalline anisotropy as a function of the Rashba parameter.
The blue circles are the simulation results and the solid line is a quadratic
fitting result. $J=1\rm~eV$, $m_e=9.1\times10^{-31}\rm~kg$, and $a=0.3\rm~nm$.
The area of the two-dimensional system in this simulation is $L\times L$ where
$L=60\rm~nm$ is the length of each direction.}
\label{Fig:TB-numerical}
\end{figure}

We consider the tight-binding Rashba model for a square lattice. From Eq.~(\ref{Eq:TB Hamiltonian}), we use the discrete crystal symmetry and
the Bloch theorem. We define
$C_{\vec{k},\sigma}=(1/\sqrt{N})\sum_{p,q}\exp(ik_xpa+ik_yqa)C_{p,q,\sigma}$,
where $N$ is the total number of sites and $\vec{k}$=$(k_x,k_y)$ is the crystal momentum within the Brillouin zone in Fig.~\ref{Fig:BZ}, which diagonalizes the Hamiltonian. We define the reduced $2\times2$ Hamiltonian $h(\vec{k})$ by $H=\sum_{\vec{k},\sigma,\sigma'}C_{\vec{k},\sigma'}^\dagger
[h(\vec{k})]_{\sigma',\sigma}C_{\vec{k},\sigma}$, where $[h(\vec{k})]_{\sigma',\sigma}$ is the matrix element of $h(\vec{k})$ in the $2\times2$ spin space. Since $h(\vec{k})$ is a $2\times2$ matrix, we compute the eigenvalues
exactly.
\begin{align}
E_\pm(\vec{k})&=-\frac{\hbar^2}{m_ea^2}(\cos k_xa+\cos k_ya)\pm\left[J^2m_z^2\phantom{\frac{\alpha_R}{a}}\right.\nonumber\\
&\quad\left.+\left(J m_x+\frac{\alpha_R}{a}\sin k_ya\right)^2+\left(J m_y-\frac{\alpha_R}{a}\sin k_xa\right)^2\right]^{1/2}.\label{Eq:energy dispersion - tight binding}
\end{align}
We plot Eq.~(\ref{Eq:energy dispersion - tight binding}) as a function of $k_y$ for $k_x=0$ in Fig.~\ref{Fig:TB-numerical}(a). The formalism given in Eq.~(\ref{Eq:fixed electron density}) provides a way to compute the magnetocrystalline anisotropy. In this section, we present the results for a two-dimensional square lattice only. The result for a two-dimensional hexagonal lattice is presented in Appendix~\ref{Sec(A):hexagonal}.

Figure~\ref{Fig:TB-numerical}(b) shows the relation between the magnetocrystalline anisotropy and the electron density (normalized to one when both majority and minority bands are completely filled). For low electron density ($N_e\lesssim0.25N_{\rm max}$), the system acquires in-plane magnetic anisotropy. This is understandable in that a parabolic approximation of the dispersion relation [Eq.~(\ref{Eq:energy dispersion - tight binding})] is equivalent to that of the free-electron Rashba model [Eq.~(\ref{Eq:quadratic E_pm})]. However, as the electron density increases, the parabolic approximation breaks down, thus the system can acquire perpendicular magnetic anisotropy from the point where the effective mass becomes negative ($N_e\approx0.25N_{\rm max}$). After this point, the perpendicular magnetic anisotropy persist widely, until $N_e\approx0.75N_{\rm max}$, covering the whole regime where the two spin bands overlap, which is in distinct contrast to the prediction [Eq.~(\ref{Eq:absence MCA})] of the idealized free-electron Rashba model.

Our computation shows a similar behavior to a first-principles
calculation~\cite{Daalderop92PRL} on the band filling dependence of the
magnetocrystalline anisotropy. Although a simple Rashba model cannot be
exact, it provides much insight into the system. Changing the electron density by means of substituting atoms or an external voltage can change not only the magnitude of the magnetocrystalline anisotropy but also its sign.

There are two key differences between the tight-binding Rashba model and the free-electron Rashba model that give rise to finite perpendicular magnetic anisotropy. The first difference is the deviation of the dispersion from a quadratic. It allows a nonzero magnetocrystalline anisotropy for a wide range of band filling, due to breakdown of Eq.~(\ref{Eq:Ne vs EF}). Once the relation between $N_e$ and $E_F$ has a magnetization dependence, a finite magnetocrystalline anisotropy can arise even if both bands are occupied. The second difference is finiteness of band width (or Brillouin zone). It plays a crucial role for the sign of the magnetocrystalline anisotropy. Since the band width is finite, there must be both maximum (band top) and minimum (band bottom) energies. Near the band bottom (the $\mathrm{\Gamma}$ point in Fig.~\ref{Fig:BZ}), the dispersion is electron-like with a positive effective mass. Thus, the theory in Sec.~\ref{Sec:Result-A} is relevant, and the sign of the magnetocrystalline anisotropy corresponds to in-plane magnetic anisotropy for low electron density. On the other hand, near the band top (the $\mathrm{M}$ point in Fig.~\ref{Fig:BZ}), the dispersion is holelike with a negative effective mass. Since the behavior is opposite to the electron-like part, the sign of the magnetocrystalline anisotropy can correspond to perpendicular magnetic anisotropy. As a result, the magnetocrystalline anisotropy near the band top of the majority band corresponds to perpendicular magnetic anisotropy [Fig.~\ref{Fig:TB-numerical}(b)]. We remark that our analysis is similar to that in \ocite{Daalderop94PRB}, which implies that most important qualitative features of the anisotropy energy can be understood by analyzing high symmetry points, where band maximum and minimum are located.

Figure~\ref{Fig:TB-numerical}(c) indicates that the magnetocrystalline
anisotropy is proportional to $\alpha_R^2$ in a reasonable range of
$\alpha_R$. We argue analytically in Sec.~\ref{Sec:Result-A} that the
magnetocryatalline anisotropy can be expanded in terms of $\alpha_R^2$. The same argument applies to this tight-binding Rashba model. We discuss below in Sec.~\ref{Sec:Result-C} the implication of the sign independence on experimental observation of the correlation between the magnetocrystalline anisotropy and other spin-orbit coupling phenomena.

\begin{figure}
\includegraphics[width=7.3cm]{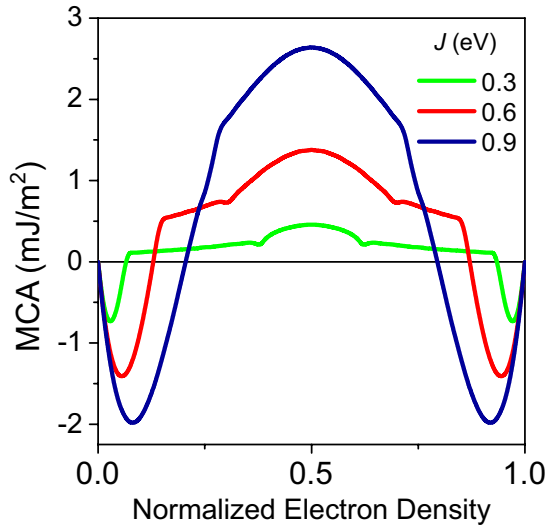}
\caption{(color online) The magnetocrystalline anisotropy (MCA) for various the
exchange energies $J=\rm0.3~eV,~0.6~eV$, and 0.9 eV with a
fixed Rashba parameter $\alpha_R=0.05\rm~eV\cdot$nm.
$m_e=9.1\times10^{-31}$ kg, and $a=0.3$ nm are used.
The stronger the exchange energy is, the higher the magnetocrystalline anisotropy is.
On the other hand, the weaker the exchange
energy is, the wider the range of the electron density that acquires perpendicular
magnetic anisotropy.
}
\label{Fig:MCA_Jsd}
\end{figure}

We now fix the Rashba parameter and compute the magnetocrystalline anisotropy for various exchange strengths. Figure~\ref{Fig:MCA_Jsd} shows the result.
The general behaviors discussed above remain the same. The weaker $J$ is, the wider the range of the emergence of perpendicular magnetic anisotropy is.
This is because the band overlap between the majority and minority bands
increases as $J$ decreases. On the other hand, the stronger $J$ is, the
stronger the magnetocrystalline anisotropy is. Therefore, we conclude that
materials with strong $J$ are advantageous to achieve a strong magnetocrystalline anisotropy with high controllability under an external voltage. On the other hand, materials with weak $J$ are advantageous for perpendicular magnetic anisotropies that stably exists over a wide range of the electron density.

The mirror symmetry of the magnetocrystalline anisotropy in
Fig.~\ref{Fig:TB-numerical}(b) originates from the symmetry between electrons at the $\rm \Gamma$ point and holes at the $\rm M$ point. From
Eq.~(\ref{Eq:energy dispersion - tight binding}), the total energy density for completely filled bands is $E_{\rm filled}=(2\pi)^{-2}\int_{-\pi/a}^{\pi/a}\int_{-\pi/a}^{\pi/a}[E_+(k_x,k_y)+E_-(k_x,k_y)]d^2k=0$,
thus the magnetocrystalline anisotropy at high electron density can be
computed by hole contributions near the $\rm M$ point. In other words,
$\Delta E(N_{\rm max}-N_e)$ is the same as the contribution from $N_e$ number of holes. Equation~(\ref{Eq:energy dispersion - tight binding}) shows the
symmetry between the electron-like $\rm \Gamma$ point and holelike $\rm M$
point, $E_\pm(k_x,k_y)=-E_\mp(\pi/a-k_x,\pi/a-k_y)$, which implies $\Delta
E(N_{\rm max}-N_e)=\Delta E(N_e)$. This is a model-specific property. For
instance, in Appendix~\ref{Sec(A):hexagonal}, we start from a two-dimensional hexagonal lattice for which the dispersion does not have such symmetry [Eq.~(\ref{Eq(A):hexagonal dispersion})] and
shows that this mirror symmetry around $N_e=0.5N_{\rm max}$ is not general.

There are four kinks in the magnetocrystalline anisotropy in
Fig.~\ref{Fig:TB-numerical}(b). We observe that the two kinks around
$N_e\approx 0.3N_{\rm max}$ and $N_e\approx 0.7N_{\rm max}$ correspond to the bottom of the minority band and the top of the majority band, respectively.
Since the minority band starts to be occupied from $N_e\approx 0.3N_{\rm
max}$, the behaviors of the magnetocrystalline anisotropy below and above
this value are different. Similarly, the majority band is no longer occupied
above $N_e\approx 0.7N_{\rm max}$. There are two more kinks near $N_e\approx
0.25N_{\rm max}$ and $N_e\approx 0.75N_{\rm max}$. We see that these occur
near the point where each band are half filled. Near these points, electrons at the Fermi level is near inflection points of the energy dispersion so the effective mass changes its sign. The existence of kinks is quite general as presented in Fig.~\ref{Fig:MCA_Jsd} and Appendix~\ref{Sec(A):hexagonal}.

To summarize this section, we perform tight-binding calculations for the
magnetocrystalline anisotropy within a discretized Rashba model. The deviation from a quadratic dispersion allows a nonzero magnetocrystalline anisotropy even when both bands are occupied. The finite band width allows emergence of perpendicular magnetic anisotropy over a wide range of the total electron density. The resulting magnetocrystalline anisotropy is proportional to $\alpha_R^2$ for a reasonable range of $\alpha_R$. Even though $\alpha_R$ becomes larger than that, the magnetocrystalline anisotropy is independent of the sign of $\alpha_R$ due to symmetry, and is constrained by symmetry to be even powers of $\alpha_R$. The implications of the sign independence and comparison with experiments are discussed in the next section. We perform similar calculations for a two-dimensional hexagonal lattice as well as a square lattice discussed here. The results are present in Appendix~\ref{Sec(A):hexagonal}.

\subsection{Remarks\label{Sec:Result-C}}

The dependence of the magnetocrystalline anisotropy on $\alpha_R$ differs from the corresponding dependence of many other phenomena of spin-orbit coupling origin. In the previous sections, we show by symmetry that the magnetocrystalline anisotropy is independent of the sign of $\alpha_R$. As a result, it is quadratic in $\alpha_R$ for a reasonable range of $\alpha_R$. On the other hand, other phenomena of spin-orbit origin such as spin-orbit torque and Dzyaloshinskii-Moriya interaction have a linear contribution in $\alpha_R$.

This feature has clear experimental implications. When a magnetic layer has two interfaces with opposite Rashba parameters, the total spin-orbit torques and the total Dzyaloshinskii-Moriya interaction arising from the both interfaces are zero since they are odd in $\alpha_R$ and the contributions from the two interfaces mutually cancel each other. However, such cancellation does not occur for the magnetocrystalline anisotropy and the contributions from the two interfaces add up since the anisotropy is even in $\alpha_R$. A similar phenomenon persists even when only one interface of a magnetic layer is subject to strong inversion asymmetry, if there are multiple energy bands. It is demonstrated~\cite{Park13PRB} that multiple bands for a given interface may experience different signs of the Rashba spin-orbit coupling. In such a situation, it is possible that contributions of those bands to the magnetocrystalline anisotropy can add up whereas their contributions to the linear spin-orbit phenomena such as spin-orbit torque and Dzyaloshinskii-Moriya interaction tend to cancel out. This observation indicates that simple proportionality analysis in experiments may fail to capture the correlation between the magnetocrystalline anisotropy and other phenomena of spin-orbit coupling origin.

In this sense, our observation can be consistent with a recent experiment~\cite{Kim16APL} reporting the opposite behaviors of the Dzyaloshinskii-Moriya interaction and the perpendicular magnetic anisotropy in Ir/Co/AlO$_x$ multilayers for various thickness of Co. According to the work, the Dzyaloshinskii-Moriya interaction reduces as the thickness of Co increases, while the perpendicular magnetic anisotropy increases. This difference may originate from multiple origins of the spin-orbit coupling phenomena, such as multiple interfaces and multiple orbital bands. As the thickness of Co increases, the contributions to the Dzyaloshinskii-Moriya interaction may cancel out while those to the magnetocrystalline anisotropy should add up. One remark is in order. Although our theory demonstrate that the positive correlation between the magnetocrystalline anisotropy and other spin-orbit coupling phenomena may breakdown, it is not necessarily the explanation of the breakdown observed in \ocite{Kim16APL} because there are other sources of magnetocrystalline anisotropy.

We observe that the magnetocrystalline anisotropy depends on the total electron density [Fig.~\ref{Fig:TB-numerical}(b)] and it can even change its sign. The strong dependence of magnetocrystalline anisotropy on the total electron density is another feature that requires a well-controlled experiment to observe the correlation. When one varies the experimental conditions to obtain systems with various spin-orbit coupling parameters, the total electron density at the interface may change, which disturbs clear interpretation of the dependence of the magnetocrystalline anisotropy on the spin-orbit coupling parameter.

The density-dependent magnetocrystalline anisotropy opens another route of the voltage-controlled magnetic
anisotropy~\cite{Nakamura10PRB,Weisheit07Science,Duan08PRL,Maruyama09NN,Wang12NM,Shiota13APL}.
The voltage-controlled magnetic anisotropy received considerable attention
due to its significant potential to enhance the efficiency of spintronic
devices. There are previous theories~\cite{Xu12JAP,Barnes14SR} suggesting
that modulating the Rashba parameter by applying an external voltage is a
possible route of the voltage-controlled magnetic anisotropy. However, it is
unlikely to be a main mechanism in metallic ferromagnetic films in which a nominal potential gradient is not a main mechanism generating Rashba
parameters~\cite{Park13PRB}. An external electric field is shielded by electron screening in the metal, thus it is difficult to change the Rashba parameter significantly. On the other hand, density variations by doping or an external voltage can change the electron density at the interface, changing the interfacial contributions to the magnetocrystalline anisotropy significantly. The conclusion from the simple model is consistent with first-principle studies~\cite{Duan08PRL,Kyuno96JPSJ}.

\section{Conclusion\label{Sec:Conclusion}}

In conclusion, we compute the magnetoctrystalline anisotropy for simple ferromagnetic Rashba models. The properties dramatically change depending on the dispersion relations. For a free electron (quadratic) dispersion, the system does not acquire
perpendicular magnetic anisotropy at all. More interestingly, we analytically show that the magnetocrystalline anisotropy is \emph{exactly} zero regardless of the Rashba coupling strength if both majority and minority bands are partially occupied in the ground state. This result is not consistent with experimental observations, implying that a free electron dispersion is not suitable for studying perpendicular magnetic anisotropy arising from the Rashba interaction.

We thus generalize the model to have a finite band width, which necessarily generates deviation from the free electron dispersion. We start from tight-binding Hamiltonians and conclude that the system acquires perpendicular magnetic anisotropy over wide range of parameters, consistent with experimental observations. A finite band width is a crucial feature of the tight-binding Hamiltonians that gives rise to perpendicular magnetic anisotropy. We also observe that the magnetocrystalline anisotropy depends on the band filling and it can even change its sign. We argue that the interface electron density modulation by voltage is a more important cause of voltage-controlled magnetic anisotropy than the voltage-controlled modulation of the Rashba parameter is.

Our results show the possibility of breakdown of positive correlation between perpendicular magnetic anisotropy and other spin-orbit coupling phenomena. In particular, if there are multiple sources of spin-orbit coupling phenomena, such as multiple interfaces and multiple orbital bands, experimental observation of the correlation requires careful analysis.

\begin{acknowledgments}
The authors acknowledge R. D. McMichael, C.-Y. You, P. M. Haney, and J. McClelland for critical reading of the manuscript. K.W.K. acknowledges support under the Cooperative Research Agreement between
the University of Maryland and the National Institute of Standards and
Technology, Center for Nanoscale Science and Technology, Grant No.
70NANB10H193, through the University of Maryland. K.W.K. was also supported
by Center for Nanoscale Science and Technology, National Institute of
Standards and Technology, based on a Collaborative Research Agreement with
Basic Science Research Institute, Pohang University of Science and
Technology. K.J.L was supported by the National Research Foundation of Korea (Grant No. 2011-028163, 2015M3D1A1070465). H.W.L. was supported by the National Research Foundation of Korea (Grant No. 2013R1A2A2A05006237) and the Ministry of Trade, Industry and Energy of Korea (Grant No. 10044723).
\end{acknowledgments}

\begin{appendix}
\section{Tight-binding Rashba model for a two-dimensional hexagonal lattice\label{Sec(A):hexagonal}}

\begin{figure}
\includegraphics[width=7.3cm]{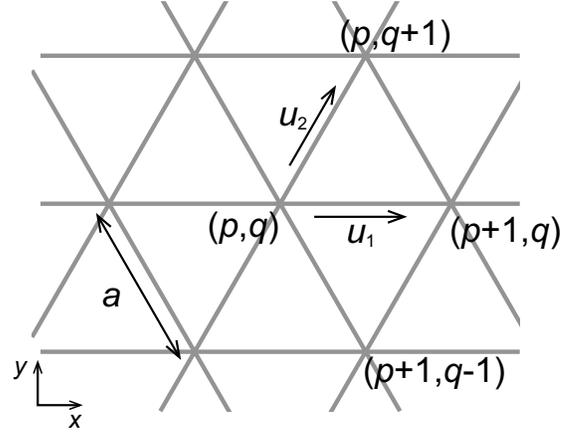}
\caption{The model for a two-dimensional hexagonal lattice. Here $a$ is the lattice constant,
$p$ and $q$ are the position indices. There are two principal directions $\vhat{u}_1$ and
$\vhat{u}_2$. Then, the position vector of each site is $pa\vhat{u}_1+qa\vhat{u}_2$.
The labeled sites by $(p,q+1)$, $(p+1,q)$, and $(p+1,q-1)$ are the neighboring hopping sites.
The other three sites are captured by adding hermitian conjugates of these.
}
\label{Fig(A):hexagonal}
\end{figure}

The two-dimensional hexagonal lattice we use here is presented in
Fig.~\ref{Fig(A):hexagonal}. We construct a tight-binding Hamiltonian by the
same way illustrated in Sec.~\ref{Sec:Model}. First we define the electron
annihilation operator $C_{p,q,\sigma}$ at the site $(p,q)$ and spin $\sigma$. The indices of a site are defined by assigning its position to be $pa\vhat{u}_1+qa\vhat{u}_2$ where $\vhat{u}=\vhat{x}$ and $\vhat{u}_2=(1/2)\vhat{x}+(\sqrt{3}/2)\vhat{y}$.
Then, the Hamiltonian is
\begin{subequations}
\begin{equation}
H=H_K+H_J+H_R,
\end{equation}
where
\begin{align}
H_K&=-t_k\sum_{pq\sigma}\left(C_{p+1,q,\sigma}^\dagger C_{p,q,\sigma}+C_{p,q+1,\sigma}^\dagger C_{p,q,\sigma}\right.\nonumber\\
&\phantom{=t_k\sum_{pq\sigma}~}\quad\left.+C_{p+1,q-1,\sigma}^\dagger C_{p,q,\sigma}\right)+\rm h.c.,\\
H_J&=J\sum_{pq\sigma\sigma'}\left[C_{p,q,\sigma}^\dagger(\vec{\sigma})_{\sigma,\sigma'}C_{p,q,\sigma'}\right]\cdot\vec{m},\\
H_R&=it_R\sum_{pq\sigma\sigma'}\left[C_{p+1,q,\sigma}^\dagger[(\vhat{u}_1\times\vhat{z})\cdot\vec{\sigma}]C_{p,q,\sigma}\right.\nonumber\\
&\phantom{it_R\sum_{pq\sigma\sigma'}(}\quad+C_{p,q+1,\sigma}^\dagger[(\vhat{u}_2\times\vhat{z})\cdot\vec{\sigma}]C_{p,q,\sigma}\nonumber\\
&\phantom{it_R\sum_{pq\sigma\sigma'}(}\quad\left.+C_{p+1,q-1,\sigma}^\dagger[((\vhat{u}_1-\vhat{u}_2)\times\vhat{z})\cdot\vec{\sigma}]C_{p,q,\sigma}\right]+\rm h.c.,
\end{align}
\end{subequations}
where $t_k$ and $t_R$ are hopping parameters to be determined.

By using Bloch theorem, the Hamiltonian can be written by 2$\times$2 matrix,
of which the eigenvalues are exactly given.
\begin{align}
E_\pm(\vec{k})&=-2t_k\left(\cos k_xa+2\cos \frac{k_xa}{2}\cos\frac{\sqrt{3}k_ya}{2}\right)\pm\tilde{J},\\
\tilde{J}^2&=J^2 m_z^2+\left(J m_x+2\sqrt{3}t_R\cos\frac{k_xa}{2}\sin\frac{\sqrt{3}k_ya}{2}\right)^2\nonumber\\
&\quad+\left(J m_y-2t_R\sin\frac{k_xa}{2}\cos\frac{\sqrt{3}k_ya}{2}-2t_R\sin k_xa\right)^2.\label{Eq(A):hexagonal dispersion}
\end{align}

The next step is determining $t_k$ and $t_R$. For a continuum limit up to
$\mathcal{O}(a^2)$,
\begin{equation}
E_\pm=-6t_k+\frac{3}{2}t_ka^2k^2\pm\sqrt{J^2+6Jt_Ra(m_xk_y-m_yk_x)+9t_R^2a^2k^2}.
\end{equation}
This should coincide with the continuum dispersion Eq.~(\ref{Eq:quadratic
E_pm}) (up to a constant energy shift). Therefore, we obtain
$t_k=\hbar^2/3m_ea^2$ and $t_R=\alpha_R/3a$.

\begin{figure}
\includegraphics[width=7.6cm]{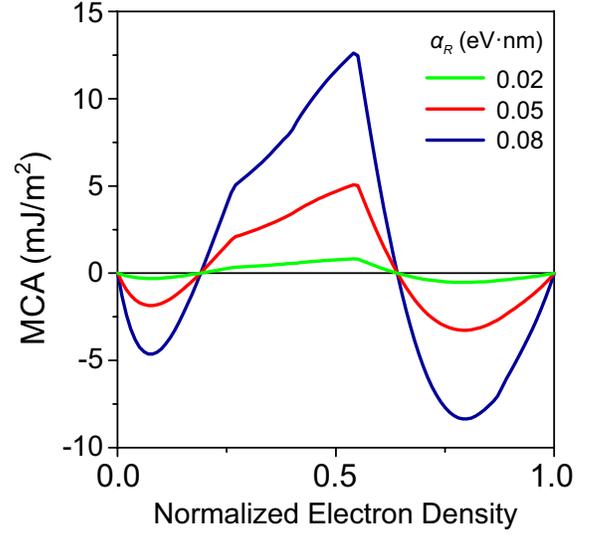}
\caption{(color online) The magnetocrystalline anisotropy computed within a tight-binding Rashba model
for a two-dimensional hexagonal lattice result (Fig.~\ref{Fig:TB-numerical}).
The overall features are the same as a two-dimensional square lattice, except the absence of the mirror symmetry
around the normalized electron density $N_e=0.5N_{\rm max}$.
}
\label{Fig(A):hexagonal_MCA}
\end{figure}

We now compute the magnetocrystalline anisotropy by the same way. The result
is shown in Fig.~\ref{Fig(A):hexagonal_MCA}. The features discussed in the main text are valid, except the model-specific property of a square lattice that the magnetocrystalline anisotropy is mirror symmetric around $N_e=0.5N_{\rm max}$.

\section{Details of the analytic theories\label{Sec:Detail}}
\subsection{Perturbation theory for a free-electron Rashba model\label{Sec:Detail-A}}
The aim of this section is to present the mathematical derivations of
Eqs.~(\ref{Eq:MCA majority only})--(\ref{Eq:Ne vs EF}) from
Eq.~(\ref{Eq:quadratic E_pm}) up to $\mathcal{O}(\alpha_R^2)$. Throughout
this section, we discard all terms beyond $\mathcal{O}(\alpha_R^2)$.

First we compute $N_\pm(E)$ from the energy dispersion. Expanding
Eq.~(\ref{Eq:quadratic E_pm}) up to $\mathcal{O}(\alpha_R^2)$, the dispersion
relation is approximated by the following quadratic function.
\begin{equation}
E_\pm(k_x,k_y)=\frac{\hbar^2k_x^2}{2\tilde{m}_\pm^x}+\frac{\hbar^2(k_y\pm k_y^0)^2}{2\tilde{m}_\pm^y}\pm J-\frac{m_e\alpha_R^2\sin^2\theta}{2\hbar^2},\label{Eq:quadratic E_pm approx}
\end{equation}
where $\theta$ is defined by $\vec{m}=(\sin\theta,0\cos\theta)$, the
spin-dependent band shift is given by $k_y^0=m_e\alpha_R\sin\theta/\hbar^2$,
and the renormalized masses are
\begin{align}
\frac{1}{\tilde{m}_\pm^x}=\frac{1}{m_e}\pm\frac{\alpha_R^2}{J\hbar^2},~\frac{1}{\tilde{m}_\pm^y}=\frac{1}{m_e}\pm\frac{\alpha_R^2\cos^2\theta}{J\hbar^2}.
\end{align}
Then, $(2\pi)^2N_\pm(E)$ is given by the area of the contour of
$E=E_\pm(k_x,k_y)$ in $\vec{k}$ space (See Fig.~\ref{Fig:electron_density}).
Since $E=E_\pm(k_x,k_y)$ forms an ellipse, the area is analytically
computable.
\begin{equation}
(2\pi)^2N_\pm(E)=\frac{2\pi\sqrt{\tilde{m}_\pm^x\tilde{m}_\pm^y}}{\hbar^2}(E-E_{\rm min}^\pm).\label{Eq:Npm vs E}
\end{equation}
From Eq.~(\ref{Eq:quadratic E_pm approx}), we obtain the band bottom energies
$E_\pm^{\rm min}$ by substituting $\vec{k}=(0,\mp k_y^0)$.
\begin{equation}
E_{\rm min}^\pm=\pm J-\frac{m_e\alpha_R^2\sin^2\theta}{2\hbar^2}.\label{Eq:Eminpm}
\end{equation}
We are now ready to compute Eq.~(\ref{Eq:total energy}).

Equation~(\ref{Eq:MCA majority only}) is derived by putting $E_F<E_{\rm
min}^+$ and $\eta=0$. Then $N_e(E)=N_-(E)$. Inverting the function, we obtain
the Fermi level as a function of the total electron density
$\varepsilon_F(N_e)=(2\pi\hbar^2/\sqrt{\tilde{m}_-^x\tilde{m}_-^y})N_e+E_{\rm
min}^-$. Equation~(\ref{Eq:total energy}) (as a function of $N_e$) is
\begin{align}
E_{\rm tot}\left(\varepsilon_F(N_e)\right)&=\int_{E_{\rm min}^-}^{\varepsilon_F(N_e)}E\frac{dN_e}{dE}dE=\int_0^{N_e}\varepsilon_F(N_e)dN_e\nonumber\\
&=\frac{\pi\hbar^2}{\sqrt{\tilde{m}_-^x\tilde{m}_-^y}}N_e^2+E_{\rm min}^- N_e.
\end{align}
Here, at the first line, we change the variable from $E$ to $N_e$ by
Eq.~(\ref{Eq:Npm vs E}). Keeping in mind that the renormalized masses and $E_{\rm min}^-$ have angular dependence, we end up with the magnetocrystalline anisotropy
\begin{align}
\Delta E(N_e)&=E_{\rm tot}\left(\varepsilon_F(N_e)\right)|_{\theta=\pi/2}-E_{\rm tot}\left(\varepsilon_F(N_e)\right)|_{\theta=0}\nonumber\\
&=-\frac{N_em_e\alpha_R^2}{2\hbar^2}\left(1-\frac{N_e}{N_-(E_{\rm min}^+)}\right),
\end{align}
where $N_-(E_{\rm min}^+)=(Jm_e/\pi\hbar^2)+\mathcal{O}(\alpha_R^2)$ from
Eqs.~(\ref{Eq:Npm vs E}) and (\ref{Eq:Eminpm}). This is Eq.~(\ref{Eq:MCA
majority only}).

To derive Eq.~(\ref{Eq:absence MCA}), we start from taking derivative of Eq.~(\ref{Eq:Npm vs E}) with respect to $E$,
\begin{equation}
\frac{dN_\pm}{dE}=\frac{\sqrt{\tilde{m}_\pm^x\tilde{m}_\pm^y}}{2\pi\hbar^2}.
\end{equation}
Therefore, Eq.~(\ref{Eq:fixed electron density}) for $\eta=1$ is given by, after some algebra,
\begin{equation}
E_{\rm tot}(E_F)=\frac{2\pi m_e(E_F^2-J^2)}{\hbar^2},\label{Eq:E_tot(E_F) both bands}
\end{equation}
which is independent of $\vec{m}$. We then combine Eq.~(\ref{Eq:Npm vs E}) with Eq.~(\ref{Eq:Eminpm}) to end up with
\begin{equation}
(2\pi)^2[N_+(E_F)+N_-(E_F)]=\frac{4\pi m_e(m_e\alpha_R^2+E_F\hbar^2)}{\hbar^4},
\end{equation}
which is nothing but Eq.~(\ref{Eq:Ne vs EF}). Inverting the function,
\begin{equation}
\varepsilon_F(N_e)=\frac{\pi \hbar^2}{2m_e}N_e-\frac{m_e\alpha_R^2}{\hbar^2}.\label{Eq:E_F both bands}
\end{equation}
Combining Eqs.~(\ref{Eq:E_tot(E_F) both bands}) and (\ref{Eq:E_F both bands}), $E_{\rm tot}(\varepsilon(N_e))$ has no angular dependence. Thus Eq.~(\ref{Eq:fixed electron density}) is
\begin{equation}
\Delta E(N_e)=E_{\rm tot}\left(\varepsilon_F(N_e)\right)|_{\theta=\pi/2}-E_{\rm tot}\left(\varepsilon_F(N_e)\right)|_{\theta=0}=0,
\end{equation}
when both bands are occupied. This is Eq.~(\ref{Eq:absence MCA}).

\subsection{Exact theory for a free-electron Rashba model\label{Sec:Detail-B}}

The purpose of this section is to show that Eq.~(\ref{Eq:absence MCA}) holds
regardless of how large $\alpha_R$ is. The flow of the proof is sketched in
Sec.~\ref{Sec:Result-A}. We first show that i) Eq.~(\ref{Eq:Ne vs EF}) is
exact above the total electron density at which both bands are occupied. This
implies that the magnetocrystalline anisotropy is independent of $E_F$ in this density range, which amounts to $E_F\ge J$. Then we show that ii) $\lim_{E_F\to\infty}\Delta E=0$. We prove this by
showing that $\Delta E$ goes $\mathcal{O}(E_F^{-1})$ at most for large $E_F$
limit. Combining i) and ii), we end up with the result that the magnetocrystalline anisotropy is exactly zero [Eq.~(\ref{Eq:absence MCA})].

\subsubsection{Proof of Eq.~(\ref{Eq:Ne vs EF}) for large $\alpha_R$}

We prove Eq.~(\ref{Eq:Ne vs EF}) by using the contour integral technique,
mainly, the Cauchy integral theorem. We do not assume that $\alpha_R$ is
small.

We assume that both bands are occupied for all $\vec{m}$. We first prove that this is equivalent to $E_F\ge J$. To show the forward part of this equivalence, we take $\vec{m}=\vhat{z}$. Then, $E_{\rm min}^+=J$, thus $E_F$ should be greater than or equal to $J$ for the minority band to be occupied. To prove the backward part, we assume $E_F\ge J$. For $\vec{k}=0$, $E_\pm(k_x,k_y)=\pm J\le J\le E_F$. Therefore, $\vec{k}=0$ state is occupied for both bands. One corollary from the proof is that $\vec{k}=0$ is always occupied when $E_F\ge J$.

We start from Eq.~(\ref{Eq:quadratic E_pm}) with $\vec{m}=(\sin\theta,0,\cos\theta)$ for $0\le\theta\le\pi/2$. We change the
variables $(k_x,k_y)$ to a single complex variable $z=i(k_x+ik_y)$. In terms
of $z$,
\begin{equation}
E_\pm(z)=\frac{\hbar^2 z^*z}{2m_e}\pm\sqrt{J^2\cos^2\theta+\alpha_R^2(z-w)(z^*-w)},
\end{equation}
where $w=J\sin\theta/\alpha_R>0$.

\begin{figure*}
\includegraphics[width=13cm]{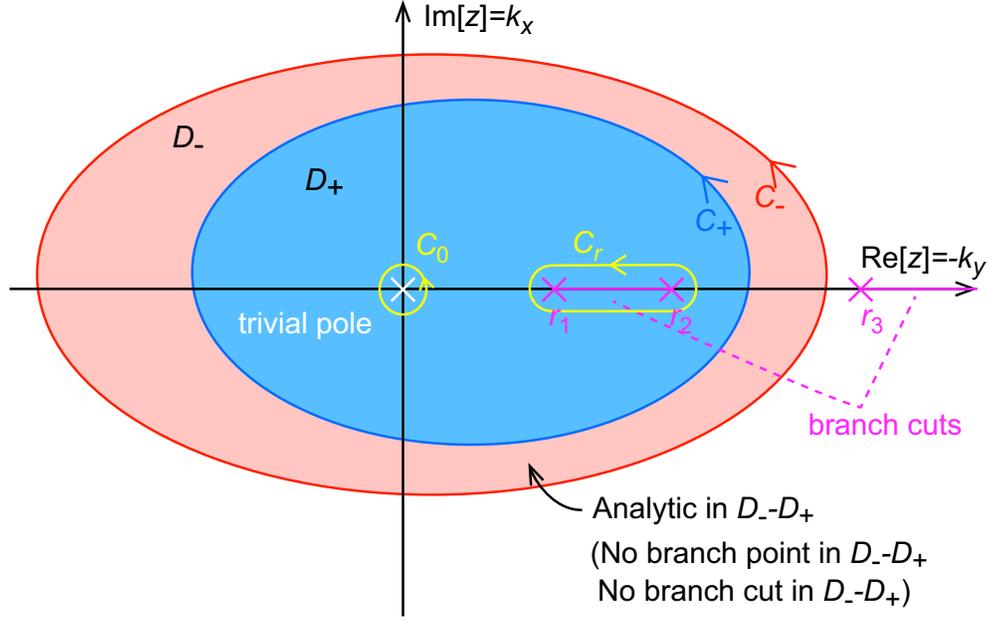}
\caption{(color online) Complex contour integral for $N_e$. Here $C_\pm$ are
the integral contours for $N_\pm$, and $D_\pm$ are the enclosed region (blue and red for $+$ an $-$) respectively. The white X is the trivial pole and the magenta Xs are the branch points of the integrand. The trivial pole is at $z=0$ and the branch points are on the real axis and denoted by $r_1$, $r_2$, and $r_3$. In Appendix~\ref{Sec(A):lemma}, we show that $r_i\in D_+$ for no or two of
$r_i$ and $r_i\notin D_-$ for the others. It is also shown that $0\in D_+$ if $E_F\ge J$. Thus, we define the branch cuts (magenta lines) by connecting $r_1$ and $r_2$, and connecting $r_3$ and a complex infinity. Therefore, the integrands in Eq.~(\ref{Eq:Ne complex expression}) are analytic in $D_--D_+$. We now can shrink the integral contour $C_\pm$ to $C_0+C_r$ (yellow) by the Cauchy integral theorem, where $C_0$ is a contour surrounding the trivial pole, and $C_r$ is a contour surrounding the branch cut defined by $r_1$ and $r_2$. If even $r_1$ and $r_2$ are not in $D_+$, $C_0$ is the only relevant contour and $C_r$ is outside $D_\pm$. Both cases give the same mathematical results.}
\label{Fig:contour}
\end{figure*}

For a given $E_F$, $N_\pm(E_F)$ is given by the area enclosed by $E_\pm(z)=E_F$ (Fig.~\ref{Fig:electron_density}). By Green's theorem, the area is given by
\begin{equation}
(2\pi)^2N_\pm=\int_{D_\pm}dk_xdk_y=\int_{C_\pm}\frac{k_xdk_y-k_ydk_x}{2}=\frac{1}{2i}\int_{C_\pm}z^*dz,
\end{equation}
where $D_\pm=\{z|E_\pm(z)\le E_F\}$ is the set of occupied states and
$C_\pm=\{z|E_\pm(z)=E_F\}$ is the boundary of $D_\pm$, that is, the contour
of the Fermi level. To perform the integration, we express $z^*$ as a
function of $z$. By equating $E_\pm(z)=E_F$ and solving $z^*$,
\begin{align}
z_{\pm~\mathrm{or}~\mp}^*&=\frac{2m_e}{z^2\hbar^4}\left[m_e\alpha_R^2(z-w)+E_F\hbar^2z\pm\sqrt{R(z)}\right],\label{Eq:zstar}\\
R(z)&=[m_e\alpha_R^2(z-w)+E_Fz\hbar^2]^2-z^2\hbar^4(E_F^2-J^2+\alpha_R^2wz).\label{Eq:Rz}
\end{align}
Here $z_\pm^*$ are functions of $z$ which satisfy $z^*=z_\pm^*(z)$ on $C_\pm$. We denote the subscript by $\pm$ or $\mp$ since it is ambiguous which one corresponds to the majority band and the minority band. However, it does not affect the final result. The total electron density is then given by
\begin{equation}
(2\pi)^2N_e(E_F)=\frac{1}{2i}\int_{C_+}z_+^*dz+\frac{1}{2i}\int_{C_-}z_-^*dz.\label{Eq:Ne complex expression}
\end{equation}

The Cauchy integral theorem implies that the complex contour integrals in
Eq.~(\ref{Eq:Ne complex expression}) is equivalent to those around
nonanalytical points only. From Eq.~(\ref{Eq:zstar}), there are two types of
nonanalytic points of $z_\pm^*$. The first one is the pole at $z=0$. We call
this the trivial pole. We show at the beginning of this section that
$(k_x,k_y)=0$ is occupied for both bands. That is, the trivial pole $z=0$ is always in $D_\pm$ (See Fig.~\ref{Fig:contour}). The second type comes from the square root function. Since the square root function is multivalued in the complex plane, there are branch cuts which connect the branch points that are defined by the zeros of $R(z)$. The whole branch cuts are nonanalytic points. Thus, it is important to see the behavior of
the zeros of $R(z)$. Since $R(z)$ is a cubic polynomial, there are three
zeros of $R(z)$. Below we present three properties of the three zeros without proofs. The proofs are presented in Appendix~\ref{Sec(A):lemma}.

The first property is that i) \emph{all three zeros of
$R(z)$ are real and nonnegative if $E_F\ge J$}. We call the zeros $r_1$, $r_2$, and
$r_3$, satisfying $r_1\le r_2\le r_3$. Another important result is that ii) \emph{$r_i\in D_-$ is equivalent to
$r_i\in D_+$}. Intuitively, we may say that, if $r_i$ is inside the contour
$C_-$, it is also inside the contour $C_+$~\footnote{This statement provides
an intuitive understanding of the result, but it is technically subtle if the
contour $C_\pm$ is not simple.}. Since $D_+\subset D_-$, one direction of the
proof is obvious, but the other direction is not. The last property is that iii) \emph{no or two zeros of $R(z)$ are
in $D_\pm$ (or inside $C_\pm$)}. As a result, the situation is summarized in
Fig.~\ref{Fig:contour}. We observe that $D_--D_+$ is analytic. Therefore, when we shrink the integral contour by using the Cauchy integral theorem, we
can end up with the same contour $C_0+C_r$ for both terms in Eq.~(\ref{Eq:Ne
complex expression}).

By using the Cauchy integral theorem, both terms in Eq.~(\ref{Eq:Ne
complex expression}) share the same integral contour.
\begin{equation}
(2\pi)^2N_e(E_F)=\frac{1}{2i}\int_{C_0+C_r}(z_+^*+z_-^*)dz.
\end{equation}
If no zeros of $R(z)$ is in $D_\pm$, $C_0$ is the only relevant contour.
However, we below show that contributions from $C_r$ are cancelled out when we
add up $z_+^*$ and $z_-^*$. One remark is in order. The situation becomes complicated if any of $r_i$ is exactly on $C_\pm$. For this case, defining $C_\pm$ bypassing $r_i$ with an infinitesimally small radius does not change the result. Another resolution is using continuity of $N_e(E_F)$. Since one of $r_i$ can be exactly on $C_\pm$ only at particular values of $E_F$, we may exclude the particular points in the proof and use the continuity to get $N_e(E_F)$ for the whole domain.

The result greatly simplifies the situation. The complicated $\sqrt{R(z)}$
terms in $z_+^*$ and $z_-^*$ are cancelled out when they are added up.
\begin{align}
(2\pi)^2N_e(E_F)&=\frac{2m_e}{i\hbar^4}\int_{C_0+C_r}\frac{\alpha_R^2m_e(z-w)+E_F\hbar^2z}{z^2}dz\nonumber\\
&=\frac{4\pi m_e}{\hbar^2}\operatorname*{Res}_{z=z_{0}}\frac{\alpha_R^2m_e(z-w)+E_F\hbar^2z}{z^2}\nonumber\\
&=\frac{4\pi m_e(\alpha_R^2m_e+E_F\hbar^2)}{\hbar^4},
\end{align}
which is exactly Eq.~(\ref{Eq:Ne vs EF}). At the second line, we use the
Cauchy's residue theorem.

The importance of the assumption that both bands are occupied in this proof
is twofold. First, the condition is equivalent to $E_F\ge J$ so that the zeros of $R(z)$ satisfy the properties proven in Appendix~\ref{Sec(A):lemma}. The properties guarantee that the integrands in Eq.~(\ref{Eq:Ne complex expression}) are analytic in $D_--D_+$ so that we can shrink the integral contours for both bands to the same contour. Second and more importantly, the complicated contributions from $\pm\sqrt{R(z)}$ are cancelled out when we add up the contributions from both bands. Therefore, we can use the Cauchy's residue theorem for the trivial pole $z=0$ only.

\subsubsection{Proof of $\lim_{E_F\to\infty}\Delta E=0$}

For extremely large $E_F$, the contour of the Fermi level is simple.
Therefore, we can define Fermi momenta for each band as a function of
the azimuthal angle of the momentum. We write $\vec{k}=(k\cos\phi,k\sin\phi)$.
Then, the Fermi momentum $k_{F,\pm}$ is defined by
$E_\pm(k_F\cos\phi,k_F\sin\phi)=E_F$. For simplicity of equations, we assume
$\alpha_R>0$, but the flow of the proof is the same for general $\alpha_R$. From Eq.~(\ref{Eq:quadratic E_pm}) and by putting
$\vec{m}=(\sin\theta,0,\cos\theta)$,
\begin{align}
k_{F,\pm}&=\sqrt{\frac{2m_eE_F}{\hbar^2}}\mp\frac{m_e\alpha_R}{\hbar^2}+\sqrt{\frac{m_e}{8E_F}}\frac{m_e\alpha_R^2\mp2J\hbar^2\sin\theta\sin\phi}{\hbar^3}\nonumber\\
&\quad\mp\frac{J^2}{4\alpha_RE_F}(1-\sin^2\theta\sin^2\phi)+\mathcal{O}(E_F^{-3/2}).\label{Eq:kF large EF}
\end{align}
By using the polar coordinate, the total energy density below the Fermi sea
is
\begin{equation}
E_{\rm tot}(E_F)=\frac{1}{(2\pi)^2}\int_0^{2\pi}d\phi\left(\int_0^{k_{F,+}}kE_+dk+\int_0^{k_{F,-}}kE_-dk\right)
\end{equation}
we can expand the integrand with respect to $1/k$ and integrate term by term
since $k_{F,\pm}$ is $\mathcal{O}(E_F^{-1/2})$. After tedious algebra, we end
up with
\begin{equation}
E_{\rm tot}(E_F)=(\theta\mbox{-independent terms})+\mathcal{O}(E_F^{-1}).
\end{equation}
Therefore, $\Delta E=\mathcal{O}(E_F^{-1})$ at most, which proves that
$\lim_{E_F\to\infty}\Delta E=0$.

\section{Properties of zeros of $R(z)$\label{Sec(A):lemma}}

In this section, we prove some important properties of zeros of $R(z)$ defined by Eq.~(\ref{Eq:Rz}). Since $R(z)$ is a cubic polynomial, it has three zeros. We call these $r_i$ for $i=1,2,3$. We below show that all of $r_i$ are real. Therefore, we can denote $r_i$ by the order of its magnitude $r_1\le r_2\le r_3$. This section consists of three subsections each of which corresponds to each property that we mention in the main text.

\subsection{All of $r_i$ are real and nonnegative if $E_F\ge J$}

We write down $R(z)=az^3+bz^2+cz+d$. Then, the coefficients are
\begin{subequations}
\label{Eq(A):R coefficients}
\begin{align}
a&=-\alpha_R^2\hbar^4w<0,\\
b&=m_e^2\alpha_R^4+2m_E\alpha_R^2E_F\hbar^2+J^2\hbar^4>0,\\
c&=-2m_e\alpha_R^2w(m_e\alpha_R^2+E_F\hbar^2)<0,\\
d&=m_e^2\alpha_R^2w^2>0.
\end{align}
\end{subequations}
Zeros of a cubic polynomial $az^3+bz^2+cz+d$ are all real if and only if $\Delta=18abcd-4b^3d+b^2c^2-4ac^3-27a^2d^2$ is nonnegative. After some algebra,
\begin{align}
\tilde{\Delta}&=4(E_F^2-J^2)(\alpha^2+2\alpha E_F+J^2)^2-27 \alpha^2J^4t^2\nonumber\\
&\quad+4\alpha J^2(\alpha+E_F)[(\alpha+E_F)^2-9(E_F^2-J^2)]t,
\end{align}
where $\tilde{\Delta}=\Delta/J^2\hbar^{12}m_e^2\alpha_R^2t$, $\alpha=m_e\alpha_R^2/\hbar^2$ and $t=\cos^2\theta$. We treat $\tilde{\Delta}$ as a function of $t$. $\tilde{\Delta}(t)$ is quadratic and the domain of $t$ is $0\le t\le1$. After some algebra,
\begin{align}
\tilde{\Delta}(0)&=4(E_F^2-J^2)(\alpha^2+2E_F\alpha+J^2)^2\ge0,\\
\tilde{\Delta}(1)&=(J^2-2\alpha E_F)^2[(\alpha+2E_F)^2-4J^2]\ge0,\\
\tilde{\Delta}_{\rm ext}&=\frac{4}{27}[3E_F^2-3J^2+(\alpha+E_F)^2]^3>0,
\end{align}
if $E_F\ge J$. Here $\tilde{\Delta}_{\rm ext}$ is the extremum value of $\tilde{\Delta}(t)$ evaluated at the value $t$ satisfying $\tilde{\Delta}'(t)=0$. Since the boundary values and the extremum value are all nonnegative, $\tilde{\Delta}$ (thus $\Delta$) is nonnegative on $0\le t\le1$, proving all of $r_i$ are real.

To show $r_i\ge 0$ for all $i$, we see the signs of the coefficients in Eq.~(\ref{Eq(A):R coefficients}). It is easy to see that $R(-z)>0$ for any real and positive $z$. Therefore, $R(z)$ has no negative real zero.

\subsection{$r_i\in D_+$ is equivalent to $r_i\in D_-$}

This statement is equivalent to that any branch point of $z_\pm^*$ cannot be in $D_--D_+$. It is one of the most important properties that allows us to draw Fig.~\ref{Fig:contour}. Since $D_+\subset D_-$, $r\in D_+\Rightarrow r\in D_-$ is straightforward, but the other direction is not.

To prove this, we use the definition of $D_\pm$ that $r_i\in D_\pm$ is equivalent to $E_\pm(r_i)-E_F\le 0$. We start from the following identity.
\begin{align}
[E_+(z)-E_F][E_-(z)-E_F]&=\left(\frac{m_e\alpha_R^2}{\hbar^2}\frac{z-w}{z}+E_F-\frac{\hbar^2z^*z}{2m_e}\right)^2\nonumber\\
&\quad-\frac{R(z)}{z^2\hbar^4}.
\end{align}
Since $R(r_i)=0$, the second term in the right-hand side is zero when $z=r_i$. In addition, we show that $r_i$ should be real in the previous section. Therefore, the first term in the right-hand side is nonnegative when $z=r_i$.
\begin{equation}
[E_+(r_i)-E_F][E_-(r_i)-E_F]\ge 0.\label{Eq(A):E+-EF E--EF}
\end{equation}
In the main text, we exclude the case where any $r_i$ is exactly on $C_\pm$. Thus, we may assume $E_\pm(r_i)-E_F\ne 0$. Under this assumption, Eq.~(\ref{Eq(A):E+-EF E--EF}) implies that $E_+(r_i)<E_F$ is equivalent to $E_-(r_i)<E_F$. In other words, $r_i\in D_+$ is equivalent to $r_i\in D_-$ for any $r_i$ satisfying $R(r_i)=0$.

\subsection{Only even number of $r_i$ are in $D_\pm$}

\begin{figure}
\includegraphics[width=7.6cm]{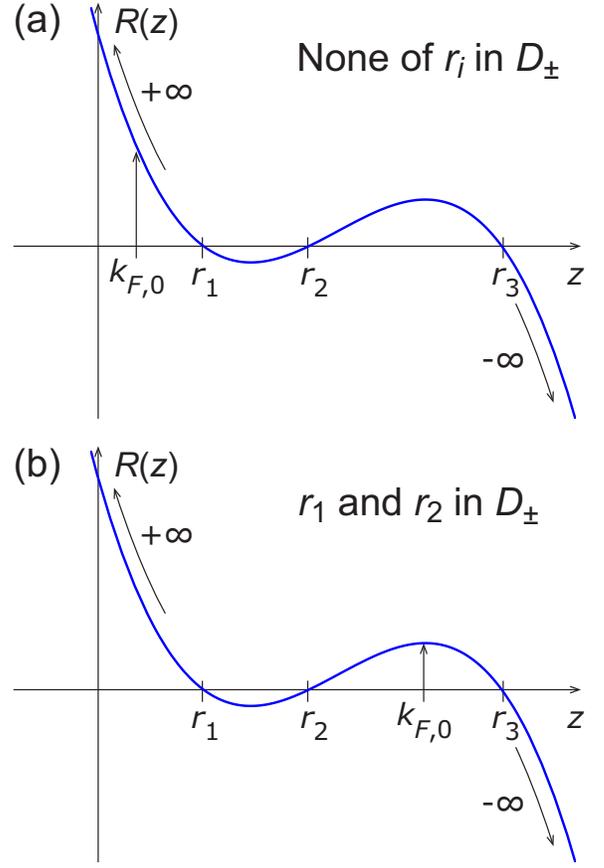}
\caption{(color online) Two possibilities of the number of $r_i$ in $D_\pm$ satisfying Eq.~(\ref{Eq:Rz condition}). Since $R(k_{F,0})>0$, the only possible domains in which $k_{F,0}$ can be present are (a) $k_{F,0}<r_1$ and (b) $r_2<k_{F,0}<r_3$. The number of $r_i$ less than $k_{F,0}$ is the number of $r_i$ in $D_\pm$. Therefore, either no or two of $r_i$ are in $D_\pm$, leading us to Fig.~\ref{Fig:contour}.
}
\label{Fig(A):Rz_roots}
\end{figure}

In the previous section, we show that the branch points of the integral Eq.~(\ref{Eq:Ne complex expression}) ($r_i$) are not in $D_--D_+$. What is important is not only the branch points but also the branch cuts. The branch cuts are defined by connecting a pair of branch points (including the complex infinity if the number of branch points are odd). To show that any branch cut does not have an intersection with $D_--D_+$, only even number of $r_i$ should be in $D_\pm$ (See Fig.~\ref{Fig:contour}).

The following lemma is useful for the proof: \emph{$r_i\in D_\pm$ is equivalent to $E_F\ge \hbar^2r_i^2/2m_e$}. This lemma is a corollary of the previous section. With this lemma, we do not need to compute $E_\pm(r_i)$ and compare to $E_F$ in order to check $r_i\in D_\pm$. Instead, we only compare $r_i$ to $\sqrt{2m_eE_F}/\hbar$~\footnote{The nonnegativity of $r_i$ plays a crucial role for deducing this.}. Therefore, it provides a useful criterion to check $r_i\in D_\pm$.

We first prove $r_i\in D_\pm\Rightarrow E_F\ge \hbar^2r_i^2/2m_e$. Since $r_i\in D_\pm$, $E_F\ge E_+(r)\ge \hbar^2r_i^2/2m_e$, which is the desired result. We next prove $E_F\ge \hbar^2r_i^2/2m_e\Rightarrow r_i\in D_\pm$. Note that $E_F\ge \hbar^2 r^2/2m_e> E_-(r_i)$, thus $r_i\in D_-$. In the previous section, we show that $r_i\in D_-$ is equivalent to $r_i\in D_+$. Therefore, $r_i\in D_\pm$, which completes the proof.

As a result, the statement that we want to show is equivalent to the statement that ``only even number of $r_i$ satisfy $r_i\le k_{F,0}$ where $k_{F,0}=\sqrt{2m_e E_F}/\hbar$." After some algebra,
\begin{equation}
R(k_{F,0})=2m_eE_F J^2\hbar^2\cos^2\theta+m_e\alpha_R^2(k_{f,0}-w)^2(m_e\alpha_R^2+2E_F\hbar^2).
\end{equation}
Therefore, $R(k_{F,0})$ is positive unless $\theta=\pi/2$ and $E_F=\hbar^2w^2/2m_e$. The latter case is not our interest because of the following argument. Note that $R(w)=\hbar^4w^2J^2\cos^2\theta$, thus, $w$ is a zero of $R(z)$ when $\theta=\pi/2$. Since $E_F=\hbar^2w^2/2m_e$, $w$ is exactly at the Fermi level (on $C_\pm$). In the main text, we exclude this situation. As a result, we now have
\begin{equation}
\lim_{z\to\mp\infty} R(z)=\pm\infty,~R(k_{F,0})>0.\label{Eq:Rz condition}
\end{equation}
Since $R(z)$ has three real and nonnegative zeros, there are only two possibilities as presented in Figs.~\ref{Fig(A):Rz_roots}(a) and \ref{Fig(A):Rz_roots}(b) respectively. Figure~\ref{Fig(A):Rz_roots} shows that either no or two of $r_i$ satisfy $r_i\le k_{F,0}$, which is the desired result.

\end{appendix}

\end{document}